\documentstyle[11pt,newpasp,twoside,epsf]{article}
\def\edcomment#1{\iffalse\marginpar{\raggedright\sl#1\/}\else\relax\fi}
\reversemarginpar

\begin{document}
\title{An Intermediate-Mass Black Hole in the Dwarf Seyfert 1 Galaxy
  POX 52}

\author{Aaron J. Barth$^1$, Luis C. Ho$^2$, and Wallace
L. W. Sargent$^1$}

\affil{$^1$~Department of Astronomy, 105-24 Caltech, Pasadena CA 91125
\\
$^2$~The Observatories of the Carnegie Institution of Washington,
  813 Santa Barbara Street, Pasadena, CA 91101}

\begin{abstract}
We describe new observations of POX 52, a previously known but nearly
forgotten example of a dwarf galaxy with an active nucleus.  While POX
52 was originally thought to be a Seyfert 2 galaxy, the new data
reveal an emission-line spectrum very similar to that of the dwarf
Seyfert 1 galaxy NGC 4395, with clear broad components to the
permitted line profiles.  The host galaxy appears to be a dwarf
elliptical; this is the only known case of a Seyfert nucleus in a
galaxy of this type.  Applying scaling relations to estimate the black
hole mass from the broad H$\beta$ linewidth and continuum luminosity,
we find $M_{\mathrm{BH}} \approx 1.6\times10^5 ~M_{\sun}$.  The
stellar velocity dispersion in the host galaxy is $36\pm5$ km
s$^{-1}$, also suggestive of a black hole mass of order $10^5
~M_{\sun}$.  Further searches for AGNs in dwarf galaxies can provide
crucial constraints on the demographics of black holes in the mass
range below $10^6 ~M_{\sun}$.
\end{abstract}

\section{Introduction}

Do dwarf galaxies host central black holes with masses below $10^6$
$M_{\sun}$?  Beyond the Local Group, dynamical detections of black
holes in this mass range are virtually impossible, but black holes
might still reveal their presence by their accretion luminosity.  Very
few examples of active galactic nuclei (AGNs) in dwarf galaxies are
known, however.  The late-type, bulgeless spiral galaxy NGC 4395 has
for several years been the only dwarf galaxy known to host a Seyfert 1
nucleus (Filippenko \& Sargent 1989). A variety of observations
suggest that its black hole has $M \approx 10^4 - 10^5 ~M_{\sun}$
(Filippenko \& Ho 2003; Shih, Iwasawa, \& Fabian 2003).

The galaxy POX 52 ($D = 93$ Mpc for $H_0 = 70$ km s$^{-1}$ Mpc
$^{-1}$) was discovered by Kunth, Sargent, \& Bothun (1987) in the POX
objective-prism survey.  They noted it as a unique example of a
Seyfert 2 nucleus in a dwarf galaxy, which they concluded was a dwarf
spiral.  Despite the unusual properties of this object, no further
follow-up observations of POX 52 were carried out since its initial
discovery.  Motivated by the possibility that POX 52 might contain a
low-mass black hole similar to the one in NGC 4395, we obtained new
optical spectra and images of POX 52 at the Keck and Las Campanas
Observatories.

\section{Observations and Results}

\begin{figure}
\plotfiddle{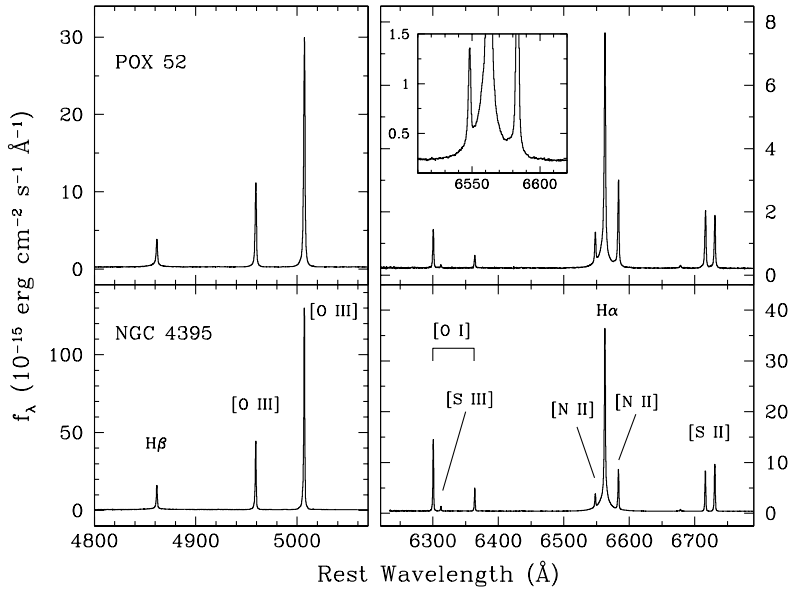}{3in}{-90}{40}{40}{-150}{240}
\caption{Keck ESI spectrum of POX 52 (upper panels) compared with NGC
  4395 (lower panels).  In the upper right panel, the inset shows a
  close-up view of the base of the H$\alpha$ emission line.}
\end{figure}

The Keck ESI spectrum shown in Figure 1 demonstrates that POX 52 has
an emission-line spectrum nearly identical to that of NGC 4395.  In
particular, POX 52 has similar broad wings on the H$\alpha$ emission
line.  Thus, POX 52 should actually be classified as a Type 1 AGN.
This is a genuine Seyfert 1 galaxy; the broad component also appears
on the higher-order Balmer emission lines as well as on He II
$\lambda4686$.  In addition, high-excitation emission lines such as
[Fe VII] are present.  Based on the widths of the emission lines, POX
52 qualifies as a narrow-line Seyfert 1 (NLS1) galaxy, although it is
an unusual member of that class since it does not have the strong Fe
II emission or the small [O~III]/H$\beta$ ratio that are typical
characteristics of NLS1s.

A virial estimate of the black hole mass can be derived from the broad
H$\beta$ linewidth and AGN continuum luminosity, using scaling
relations derived from reverberation mapping of Seyfert galaxies, and
assuming gravitational motion of the broad-line clouds.  We fit the
H$\beta$ profile using a model consisting of a broad Gaussian plus a
narrow component having the same shape as the [O III] $\lambda5007$
line.  The best-fitting model has a broad-line FWHM of 760 km
s$^{-1}$.  Combining this with the AGN continuum luminosity at 5100
\AA\ using the scaling relations from Kaspi et al.\ (2000) gives a
mass estimate of $\sim1.6\times10^5 ~M_{\sun}$.  This estimate is
highly uncertain, since it requires extrapolating the scaling
relations far beyond the mass range over which they have been
calibrated.  Nevertheless, this result suggests that POX 52 hosts a
black hole with a mass that is substantially smaller than those of
typical Seyfert galaxies.  Also, if POX 52 has a spectral energy
distribution similar to that of NGC 4395 and if it is radiating at $L
< L_{\mathrm{Edd}}$, then a lower limit to its black hole mass is $3
\times 10^4 ~M_{\sun}$.

\begin{figure}
\plottwo{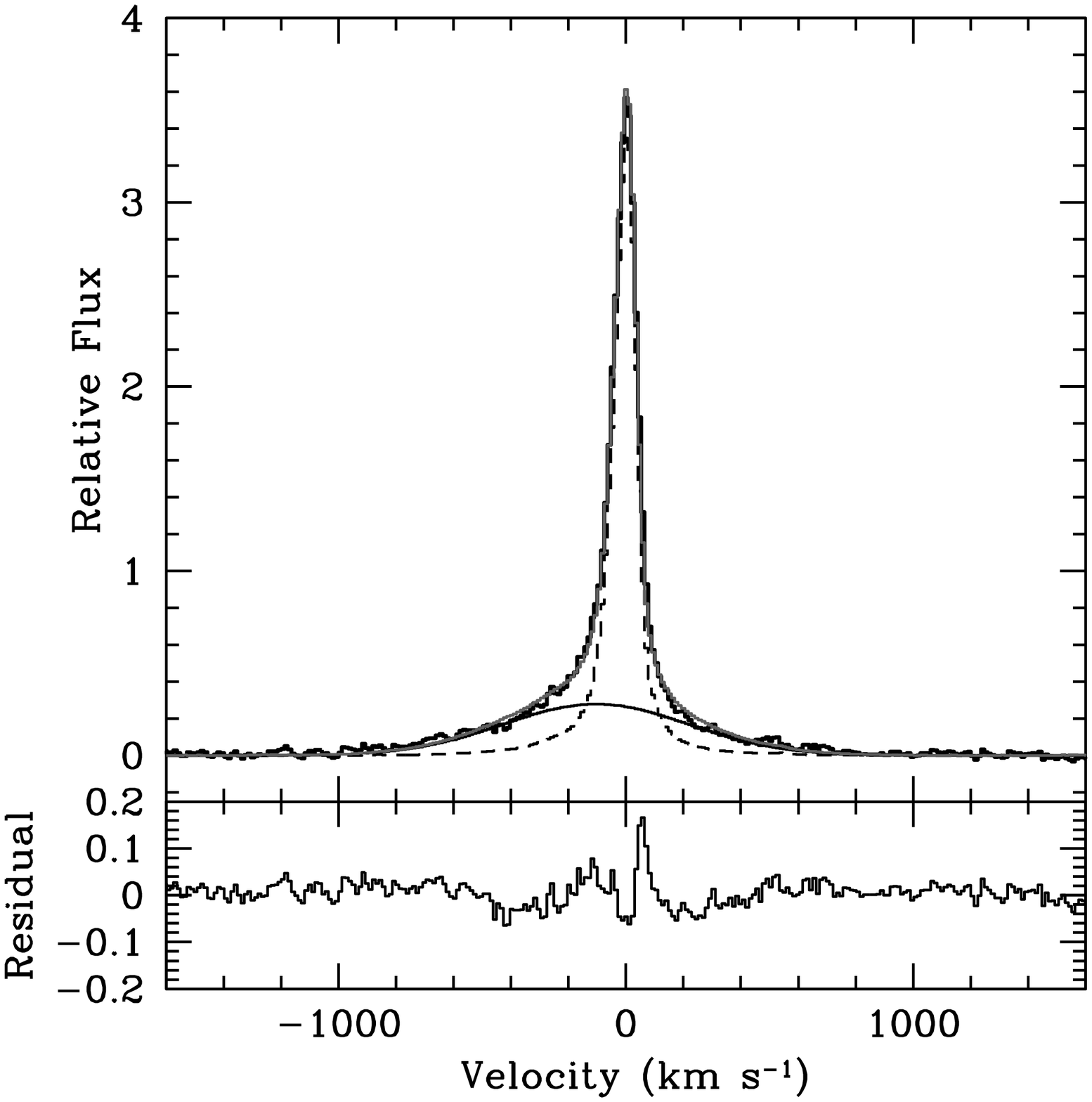}{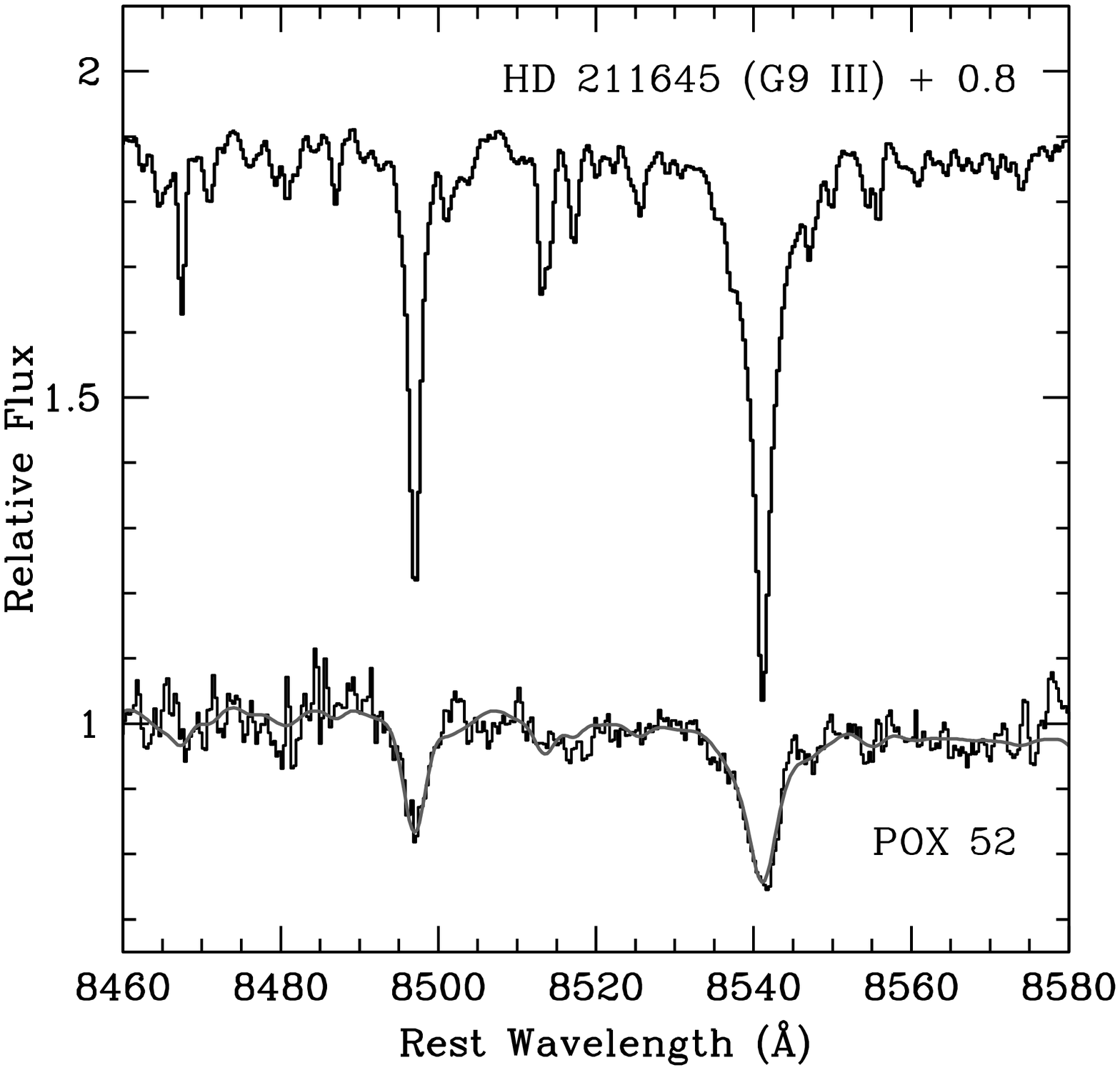}
\caption{\emph{Left panel:} Model fit to the H$\beta$ emission line.
  The dashed curve is a scaled version of the [O III] line profile,
  used to model the narrow H$\beta$ component.  \emph{Right panel:}
  The Ca II $\lambda\lambda8498, 8542$ lines in POX 52.  The model fit
  shows the G9 III template star diluted by a featureless continuum
  and broadened by $\sigma = 36$ km s$^{-1}$.}
\end{figure} 

From the Ca II triplet lines in the ESI spectrum, we find a stellar
velocity dispersion of $36 \pm 5$ km s$^{-1}$.  This is the
second-smallest velocity dispersion known for any AGN; the smallest is
NGC 4395 with $\sigma < 30$ km s$^{-1}$ (Filippenko \& Ho 2003).
Extrapolating the $M_{\mathrm{BH}}-\sigma$ relation of Tremaine et
al. (2002) to $\sigma=36$ km s$^{-1}$, the expected black hole mass is
$\sim1.3 \times 10^5 ~M_{\sun}$.  This is surprisingly close to the
mass estimate derived from the H$\beta$ linewidth.  The [O~III]
emission line has FWHM = $87 \pm 10$ km s$^{-1}$ or $\sigma = 37 \pm
4$  km s$^{-1}$, so the stellar and gaseous velocity dispersions are
nearly equal.

New \emph{BVRI} images of POX 52 were obtained at the 2.5-m du Pont
telescope at Las Campanas Observatory in 0\farcs75 seeing.  Using the
GALFIT profile-fitting code (Peng et al.\ 2002), we performed a
decomposition into point-source and host galaxy components.  An
exponential profile gives an extremely poor fit to the host galaxy,
while a S\'ersic profile with an index of $3.6 \pm 0.2$ and an effective
radius of $\sim0.5$ kpc fits the galaxy adequately.  The host galaxy
has $M_B = -16.8$ mag and \bv\ = 0.8 mag, consistent with the
properties of a dwarf elliptical galaxy.  No spiral structure or knots
suggestive of star-forming regions are detected.  In the fundamental
plane, POX 52 lies close to the Virgo dwarf elliptical galaxies
studied by Geha, Guhathakurta, \& van der Marel (2003).  Thus, it
appears that POX 52 is the first known example of a Type 1 AGN in a
dwarf elliptical galaxy.

\begin{figure}
\plottwo{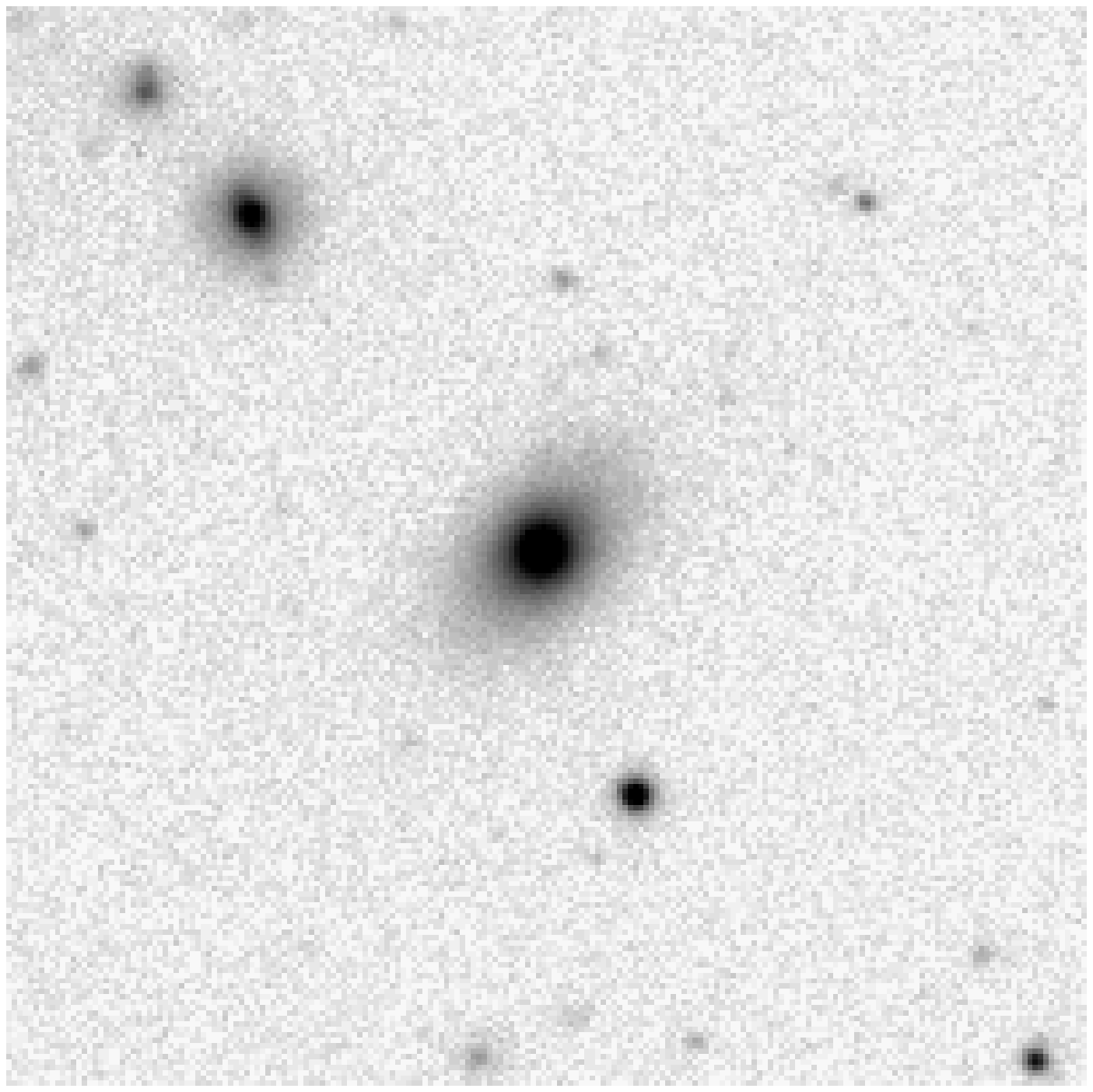}{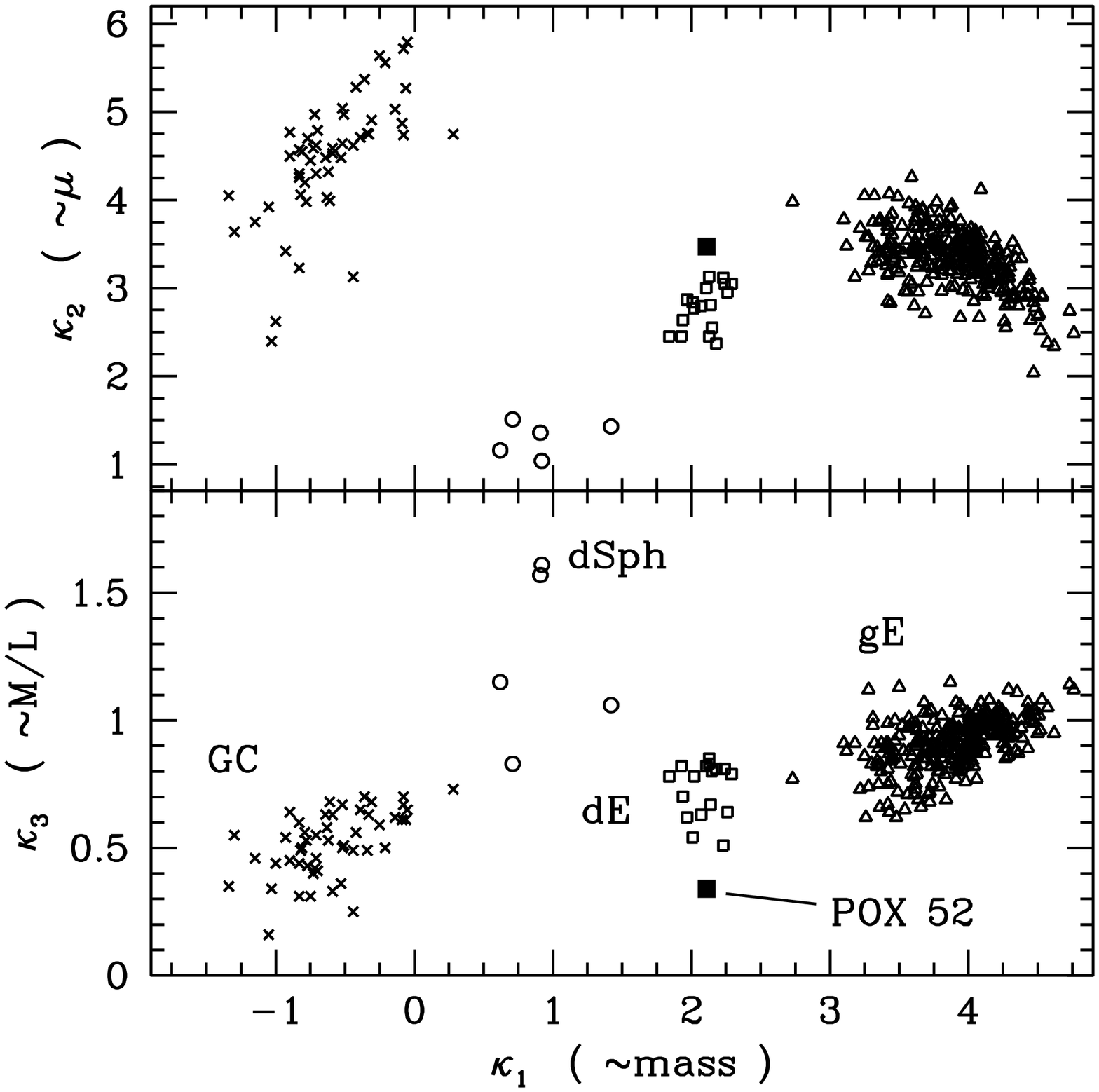}
\caption{\emph{Left panel:} $R$-band image centered on POX 52.  The
  image size is 52\arcsec\ on a side.  \emph{Right panel:}
  $\kappa$-space projections of the fundamental plane, including
  globular clusters (crosses), dwarf spheroidals (circles), giant
  ellipticals (triangles), dwarf ellipticals (open squares), and POX
  52 (filled square).  Literature data are from Burstein et al.\
  (1997) and Geha et al.\ (2003).}
\end{figure}

\section{Conclusions and Future Work}

POX 52 is one of only two dwarf galaxies known to contain an AGN, and
its black hole mass is likely to be of order $10^5 ~M_{\sun}$.  We hope
to obtain \emph{HST} imaging in the future, for a definitive
measurement of the host galaxy profile, as well as X-ray data from
\emph{Chandra} and/or \emph{XMM-Newton} to study its AGN in greater detail
and search for variability.  

Currently, we have almost no information on the population of black
holes with masses below $10^6 ~M_{\sun}$.  Any constraints on black
hole demographics in this mass range would be of particular interest
for future gravitational wave experiments.  Further searches for AGNs
in dwarf galaxies may be the best way to improve this situation, and
can yield at least a lower limit to the fraction of dwarf galaxies
(both ellipticals and spirals) that contain central black holes.  We
are currently beginning a search of the SDSS data archive to identify
additional dwarf galaxies with active nuclei.

\references

\reference{Burstein, D., Bender, R., Faber, S., \& Nolthenius,
  R. 1997, \aj, 114, 1365}

\reference{Filippenko, A. V., \& Ho, L. C. 2003, \apj, 588, L13}

\reference{Filippenko, A. V., \& Sargent, W. L. W. 1989, \apj, 342, L11}

\reference{Geha, M., Guhathakurta, P., \& van der Marel, R. 2003, \aj,
  in press}

\reference{Kaspi, S., et al. 2000, \apj, 533, 631}

\reference{Kunth, D., Sargent, W. L. W., \& Bothun, G. D. 1987, \aj,
  93, 29}

\reference{Peng, C. Y., Ho, L. C., Impey, C. D., \& Rix, H.-W.  2002,
  \aj, 124, 266}

\reference{Shih, D. C., Iwasawa, K., \& Fabian, A. C. 2003, \mnras,
  341, 973}

\reference{Tremaine, S., et al.\ 2002, \apj, 574, 740}

\end{document}